\documentclass[AMA,Times1COL]{WileyNJD-v5}

\usepackage{todonotes}
\usepackage{siunitx}
\setuptodonotes{inline}

\articletype{Article}

\received{26 April 2016}
\revised{6 June 2016}
\accepted{6 June 2016}

\begin{document}
\newcommand{\titletext}{Update on the German and Australasian Optical Ground Station Networks}
\title{\titletext\protect}

\author[1]{Nicholas J. Rattenbury*}
\author[1]{Joseph Ashby}
\author[2]{Francis Bennet}
\author[2]{Marcus Birch}
\author[3]{John E. Cater}
\author[4]{Kate Ferguson}
\author[5]{Dirk Giggenbach}
\author[6]{Ken Grant}
\author[7]{Andreas Knopp}
\author[8]{Marcus T. Knopp}
\author[9,10]{Ed Kruzins}
\author[11]{Andrew Lambert}
\author[6]{Kerry Mudge}
\author[1]{Catherine Qualtrough}
\author[5]{Samuele Raffa}
\author[5]{Jonas Rittershofer}
\author[1]{Mikhael T. Sayat}
\author[12]{Sascha Schediwy}
\author[7]{Robert T. Schwarz}
\author[13]{Matthew Sellars}
\author[6]{Oliver Thearle}
\author[2]{Tony Travouillon}
\author[1]{Kevin Walker}
\author[12]{Shane Walsh}
\author[1,14]{Stephen Weddell}

\authormark{Rattenbury \textsc{et al.}}

\address[1]{\orgdiv{Department of Physics}, \orgname{The University of Auckland}, 
\orgaddress{\state{Auckland}, \country{New Zealand}}}

\address[2]{\orgdiv{College of Science}, \orgname{Australian National University},
\orgaddress{\state{ACT}, \country{Australia}}}

\address[3]{\orgdiv{Department of Mechanical Engineering}, \orgname{University of Canterbury}, \orgaddress{\state{Christchurch}, \country{New Zealand}}}

\address[4]{\orgdiv{ANU Institute for Space}, \orgname{Australian National University}, \orgaddress{\state{ACT}, \country{Australia}}}

\address[5]{\orgdiv{Institute of Communications and Navigation}, 
\orgname{German Aerospace Center (DLR e.V.)},
\orgaddress{\state{Wessling}, \country{ Germany}}}

\address[6]{\orgdiv{Defence Science and Technology Group}, 
\orgname{Edinburgh}, 
\orgaddress{\state{SA}, \country{Australia}}}

\address[7]{\orgdiv{Space Research Center}, \orgname{University of the Bundeswehr Munich}, 
\orgaddress{\state{Neubiberg}, \country{Germany}}}

\address[8]{\orgdiv{Responsive Space Cluster Competence Center (RSC\textsuperscript{3})}, 
\orgname{German Aerospace Center (DLR e. V.)}, 
\orgaddress{\state{Wessling}, \country{Germany}}}

\address[9]{\orgdiv{Canberra Space}, \orgname{University of New South Wales}, 
\orgaddress{\state{NSW}, \country{Australia}}}

\address[10]{\orgdiv{Commonwealth Scientific Industrial Research Organisation}, 
\orgname{Sydney}, 
\orgaddress{\state{NSW}, \country{Australia}}}

\address[11]{\orgdiv{School of Engineering and Information Technology}, 
\orgname{University of New South Wales}, 
\orgaddress{\state{ACT}, \country{Australia}}}

\address[12]{\orgdiv{International Centre for Radio Astronomy Research}, 
\orgname{The University of Western Australia}, 
\orgaddress{\state{WA}, \country{Australia}}}

\address[13]{\orgdiv{Research School of Physics}, \orgname{Australian National University}, \orgaddress{\state{ACT}, \country{Australia}}}

\address[14]{\orgdiv{Department of Electrical Engineering},
\orgname{University of Canterbury},
\orgaddress{\state{Christchurch}, \country{New Zealand}}}


\abstract[Abstract]{
Networks of ground stations designed to transmit and receive at optical wavelengths through the atmosphere offer an opportunity to provide on-demand, high-bandwidth, secure communications with spacecraft in Earth orbit and beyond.
This work describes the operation and activities of current Free Space Optical Communication (FSOC) ground stations in Germany and Australasia. In Germany, FSOC facilities are located at the Oberpfaffenhofen campus of the Deutsches Zentrum für Luft- und Raumfahrt (German Aerospace Center, DLR), the Laser-Bodenstation in Trauen (Responsive Space Cluster Competence Center, DLR), and the Research Center Space of the University of the Bundeswehr Munich in Neubiberg. The DLR also operates a ground station in Almeria, Spain as part of the European Optical Nucleus Network. 
The Australasian Optical Ground Station Network (AOGSN) is a proposed network of 0.5 -- 0.7~\si{\meter} class optical telescopes located across Australia and New Zealand. The development and progress for each node of the AOGSN is reported, along with optimisation of future site locations based on cloud cover analysis. 
}


\jnlcitation{\cname{
\author{N.J. Rattenbury}, 
(\cyear{2023}), 
\ctitle{\titletext}, \cjournal{IJSCN}, \cvol{TBD}.}}

\maketitle

\section{Introduction}
\label{sec:Intro}
Free space optical communication (FSOC) is a technique that uses lasers to communicate information. These lasers are typically in the infra-red band and communication protocols can be modulated on properties such as laser amplitude or phase. Laser light offers substantial advantages for inter-satellite and Earth-satellite communications over conventional radio frequency (RF) communications~\cite{ANBARASI2017161,book2010}. Channel bandwidths available using FSOC are higher than those using conventional radio frequencies~\cite{Reyes2005}, and the transmitter and receiver hardware is smaller, lighter and more efficient~\cite{Kubo2012,Buchheim2012} while also avoiding spectrum licensing. These features are attractive for the designers of spacecraft, where the size, weight and power budgets are highly constrained. Demand for higher quantities of data from space is increasing for conventional activities such as Earth observation and global telecommunications~\cite{Bennet2020}. Ever higher rates of data will be required as crewed space missions start to take place in cislunar space, on the lunar surface and further afield to Mars~\cite{hemmati2006deep,Biswas2018}. FSOC also offers a communication channel that can be made robust against eavesdropping using quantum processes~\cite{Bennet2018,Mink2009,Bedington2017,Letzter2018}. This is a feature available with FSOC that is of particular interest to sectors where security is paramount, including government, defence and global banking.

Further experimentation and technology development is required to enable FSOC to achieve these features for the international space-based communication market. This is particularly important for high-performance systems requiring advanced techniques such as coherent communications and adaptive optics. These research and development activities will be accelerated by the development of research-grade optical ground station terminals. 

An FSOC ground station requires a cloud-free line of sight between it and a client space asset to establish a communications channel. Even in sites with the highest fractions of cloud-free skies, there will be periods that an FSOC terminal at such a site will be obscured by cloud, or unavailable for operations for other reasons such as the need to perform maintenance. A network of FSOC ground stations provides a mitigation against these situations encountered by a single FSOC ground station. Such a network would comprise a number of FSOC ground station nodes, where each node is sufficiently distant from the others so that the weather conditions at that node are as uncorrelated as possible with those experienced at the other nodes.

Cloud coverage is the principal constraint on the performance of an optical ground station (OGS) network, and predicting network performance has been the subject of a significant quantity of literature. The DLR has developed a model for network performance estimation with pre-selected sites around Europe and a means of diversity estimation using spatial correlations of cloud cover~\cite{fuchs2015ground}. Assuming site independence or empirical outage measurements, large European networks are proposed with reliability of 98\% to 99\%. The ONUBLA software was also proposed as means to use these methods via software, collating and analysing cloud fraction for any proposed OGS network, and predicts the performance of the network~\cite{onubla, fuchsONUBLA, fuchsONUBLAESA}. 

Diversity and reliability for an Australasian Optical Ground Station Network (AOGSN) was further explored in a recent publication, which also proposes a complete analytical solution to network diversity for spatially correlated OGS which did not exist in previous attempts~\cite{birch2023availability}. A base AOGSN with three nodes ($N=3$) was found to have a 6.4\% outage probability and a larger AOGSN with eight nodes ($N=8$) had an 0.02\% outage probability. A spatially-resolved means of network optimisation, i.e., without pre-selecting locations, was also shown. Australia and New Zealand were generally shown to be excellent locations for an OGS network due to the low cloud fraction and large geographic size which offers minimal site-to-site correlation and broad coverage~\cite{birch2023availability}.

In this work an overview of the establishment of optical ground stations in Europe and Australasia is presented, as well as a report on the optimisation of node placement to most efficiently expand on existing infrastructure with site optimisation in New Zealand. 

\section{German FSOC Ground Stations and Network}
\label{sec:German}
Several research institutes in Germany have established, or are establishing FSOC ground stations, and programs of FSOC research. The German Aerospace Centre (DLR), is leading research on the definition of FSOC standards and proposed global FSOC-enabled communication solutions~\cite{borowy2018satcom}. A significant step towards global connectivity has been the establishment of the European Optical Nucleus Network initiative (EONN)~\cite{krynitz2021european}. Space Agencies and industry have allied to create a multi-site, multi-mission OGSN which supports common optical space communications standards. Participating parties contribute operations time on self-funded optical ground stations to an integrated network that is made available to the space community as a service. The DLR, as one of the founding members, contributes its OGS in Almeria, Spain, to the EONN. The DLR has also developed extensive expertise on optical ground stations over the years and it is continuously extending its own optical antenna park.

\subsection{Technological Precursor Site: DLR Oberpfaffenhofen}
\label{sec:Obf}

\begin{figure}[!htb]
    \centering
    \includegraphics[width=0.5\linewidth]{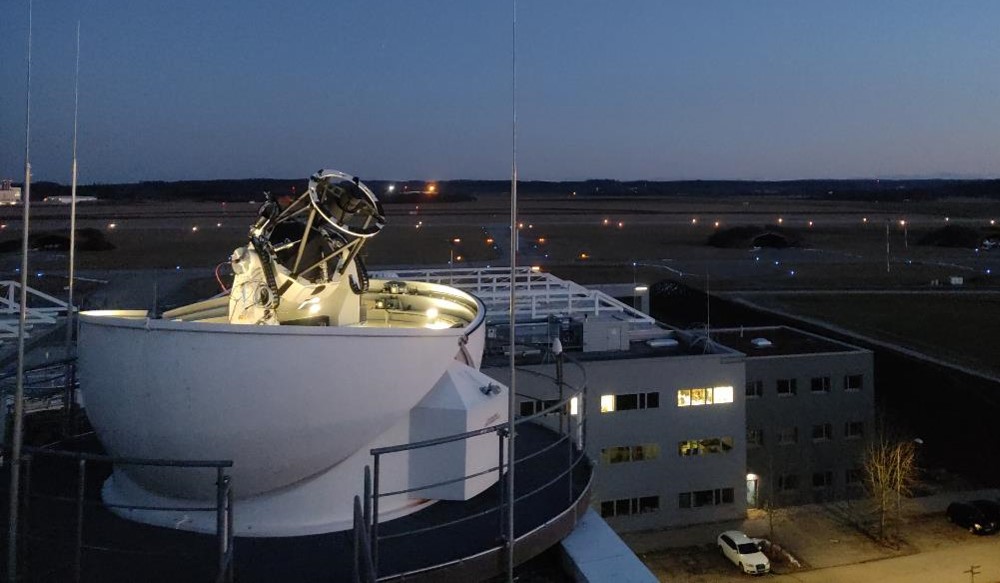}
    \includegraphics[width=0.31\linewidth]{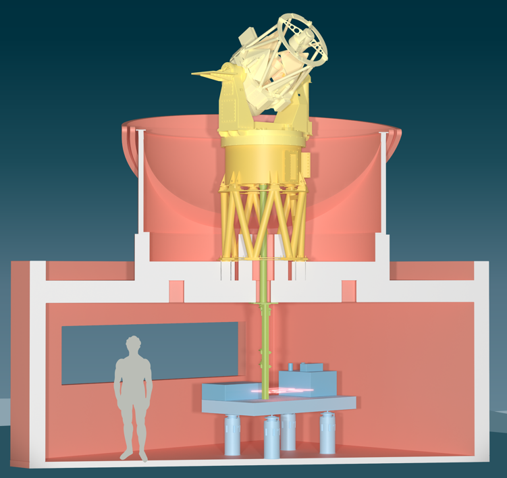}
    \caption{The OGSOP-new 80~\si{\centi\meter} Ritchey-Chretien Telescope, and adjacent functional Diagram of OGSOP-new with Coude-Room underneath hosting various experiments for AO and QKD. Image: DLR}
    \label{fig:OGSOP_new}
\end{figure}

The DLR Institute of Communications and Navigation (IKN) has developed optical ground stations since 2004, starting with a 40~\si{\centi\meter} telescope for orbital checkout of the first coherent satellite Laser Communications Terminal (LCT) onboard the TerraSAR-X Satellite, with campaigns in Oberpfaffenhofen and Calar Alto, see Figure~\ref{fig:OGSOP_new}. This ground station was later permanently installed on the rooftop of IKN for experiments with the Japanese OICETS/Kirari satellite with its LUCE-Terminal, performing the first LEO satellite data downlinks in Europe~\cite{dlr51119}. Additional  experiments were performed (such as aeronautical quantum key distribution~\cite{dlr81410}, or links with the OPALS terminal by JPL on the ISS and NICT's SOKRATES~\cite{dlr110518}), before the facility underwent a major refurbishment, including a larger telescope (80~\si{\centi\meter}) and the addition of a Coud\'{e} beam path to a laboratory room.

The OGSOP-new became operational in 2021 and is currently employed in bidirectional links with adaptive optics (AO) and AO-predistortion to geostationary satellites, and for the checkout of the various LEO satellites carrying diverse OSIRIS-terminals (Optical Space InfraRed link System)~\cite{fuchs2019update}, namely the Flying Laptop~\cite{dlr130743}, BIROS~\cite{dlr89346}, and PIXL~\cite{PIXL2023}. Future utilization will be for tests with OSIRISv3 for data downlinks with more than 10~\si{\giga\bit\per\second}, checkout of inter-satellite link terminals (CubeISL), and quantum Key Distribution (QKD) satellite terminals (QUBE)~\cite{dlr135701}.

\begin{figure}
    \centering
    \includegraphics[width=0.7\linewidth]{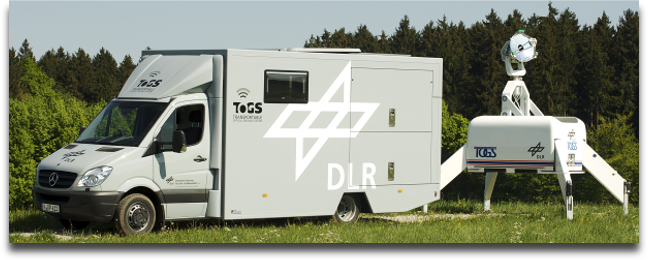}
    \caption{Transportable Optical Ground Station (TOGS) with transport-van housing also the operations room. Image: DLR}
    \label{fig:TOGS}
\end{figure}

The TOGS (Transportable Optical Ground Station, by DLR-IKN) was built in 2011 to enable temporary OGS operations at remote locations. It consists of a 60~\si{\centi\meter} aluminium telescope in an airfreight-sized box that also holds the operations computer and any receiver and transmitter equipment for data transmission, as well as mechanics to extend the telescope and mount from its  container, see figure~\ref{fig:TOGS}. This assembly can be stored and transported by a custom-built vehicle that also holds an operations room. TOGS was operated on the Island of La Palma for demonstrating the optical remote control of robotic operations from space~\cite{dlr107384}, and performed FSOC links to a hypersonic jet fighter in flight~\cite{moll2015ground}. It is now regularly used for OSIRIS checkout operations and in conjunction with other DLR-OGSs conducted joined observations of LEO LASER downlinks~\cite{giggenbach2022beambias}. 

Based on the developments for the IKN-OGSs, developments for operational ground stations in Almeria (FOGATA) and at the DLR-site Trauen (LaBoT) are in progress, making use of the SOFA unit (see Section~\ref{sec:SOFA}) as communications equipment. In preparation for these sites, the German Space Operations Center (GSOC) has realized a test facility at its premises in Oberpfaffenhofen. Residing on the rooftop of the control center building, this installation serves to host the initial engineering models of the station equipment used at the remote OGS sites. Instruments are integrated in the operational network and control environment at GSOC for verification and validation campaigns at system and sub-system level. Protected by a roll-off cabin, diverse telescopes and mounts have been installed. The latest setup consists of a mid-range COTS telescope (TS 12\"--f/8-Ritchey-Chr\'{e}tien-Astrograph, TS-Optics, Munich, Germany) at a precision mount in Alt-Az configuration (L-350, PlaneWave Instruments, Adrian, Michigan). Depending on the particular instrument attached to the telescope, the measurements include (1) pointing and tracking accuracy with a camera-based setup~\cite{giggenbach2022beambias}; (2) optical signal strength with an InGaAs-PIN-based photoreceiver (OE-200-S, FEMTO, Berlin, Germany); (3) bit-error-rates utilizing optimized detector heads (APDRFE-100M and APDRFE-1G, joint development of DLR and LiPaCom, Puergen, Germany). The setup can be equipped with the DLR SOFA to create a complete OGS including different transmit sources. All components are remotely controlled using an in-house developed monitoring and control (M\&C) system~\cite{jain2022monitoring}.

\subsection{German Optical Ground Station Network}
\label{GOGSN}

\begin{figure}[!htb]
\begin{center}
    \includegraphics[width=0.8\linewidth]{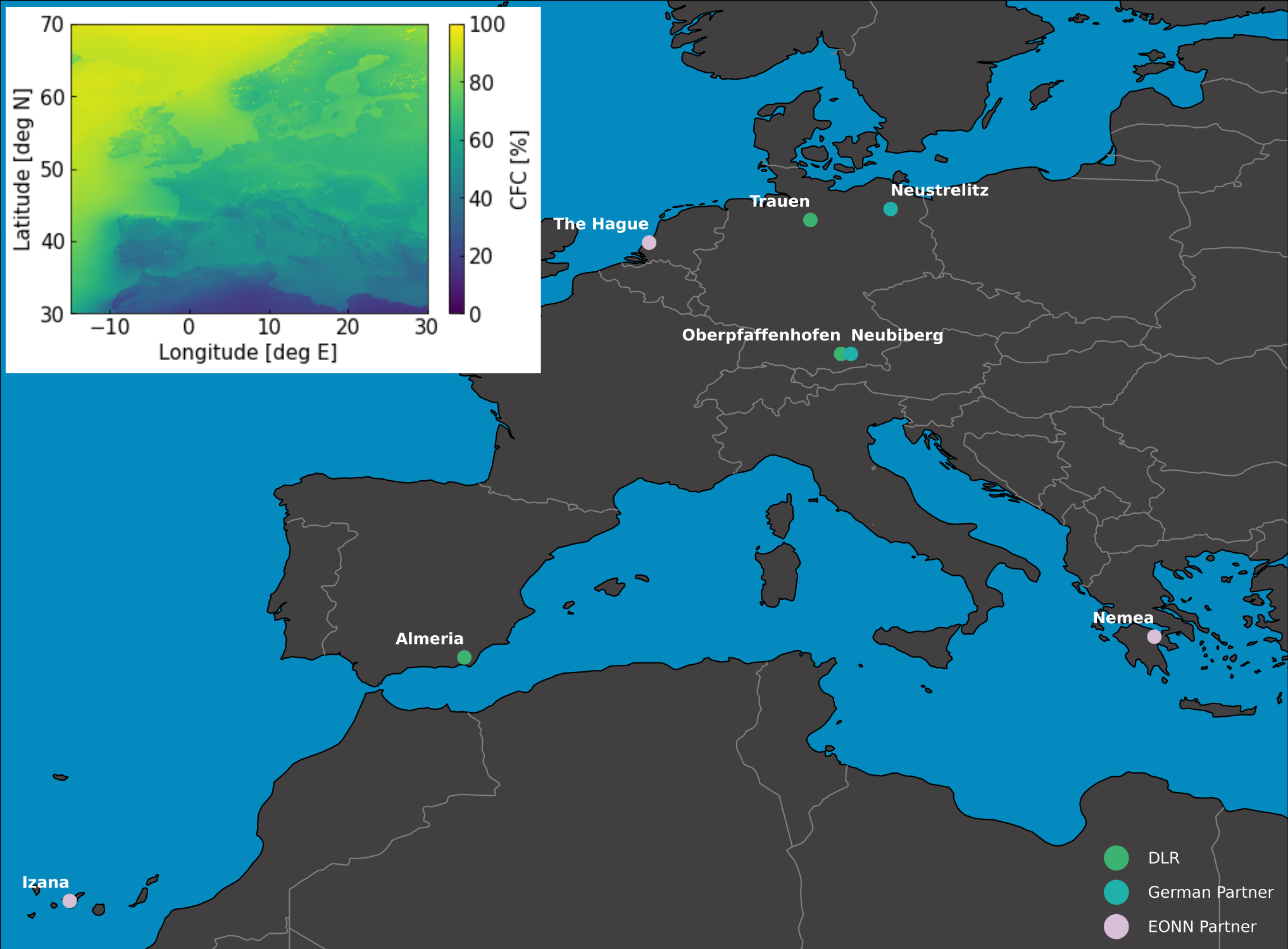}
\end{center}
    \caption{German Optical Ground Station Network with access to the European Optical Nucleus Network. \textit{Inset:} Average Cloud Fraction Coverage (CFC) in Europe\cite{finkenspier2020}.}
    \label{fig:GOGSN}
\end{figure}

The DLR manages the integration of several optical ground stations into a German OGS network (GOGSN, cf. Fig.~\ref{fig:GOGSN}): DLR/GSOC is building the Free-space Optical Ground Antenna TAbernas (FOGATA)\cite{lantschner2019optical} at the Plataforma Solar de \textbf{Almeria}, a solar research site of the Spanish Centre for Energy, Environmental and Technological Research (CIEMAT). DLR/RSC\textsuperscript{3} (Responsive Space Cluster Competence Center) together with their prime contractor DiGOS Potsdam are implementing the Laser-Bodenstation (\textit{engl: Laser Groundstation}) \textbf{Trauen} (LaBoT)~\cite{kohler2023commissioning,kohler2022labot}, the SPACE Research Center of the University of the Bundeswehr Munich is currently setting up the affiliated OGS at their \textbf{Neubiberg} (NBB) campus~\cite{schwarz2023optical} and the Technical University of Berlin is going to realize a combined FSOC and Satellite Laser Ranging (SLR) station at the DLR site \textbf{Neustrelitz} (NSG). The new Optical Ground Station Oberpfaffenhofen (OGSOP-NG) of DLR/KN is already operational and is used for various FSOC experiments~\cite{prell2023ogsop}.

Table~\ref{tab:ogsfeatures} gives an overview on the specifications of those ground stations. Network load is handled via a reactive planning process, taking into account spacecraft visibilities and terminal constraints, as well as actual weather conditions at the OGS locations~\cite{dlr197490}. Station telemetry is shared utilizing the upcoming CCSDS standard for Service Management~\cite{pietras2010} allowing for automated scheduling. In addition, the European Optical Nucleus Network can be accessed via the respective application programming interface (API). All OGSs follow the design of a robotic observatory consisting of an RC-telescope on a precision mount within a dome and with a (containerized) control section hosting the electronic system. They will be featured with a common instrument (SOFA) to harmonize their air-to-ground interface gearing them towards compatibility with the upcoming CCSDS standard O3K~\cite{edwards2022}.

\begin{table}[!htb]
    \centering
    \caption{Features of the German Optical Ground Station Network Nodes}
    \begin{tabular}{lrrrrr}
    \toprule
        & \textbf{FOGATA} & \textbf{LaBoT} & \textbf{OGS NBB} & \textbf{OGS NSG (prelim.)} & \textbf{OGSOP-NG}\\ 
        \midrule
        Location & Almeria, Spain & Trauen, Germany & Neubiberg, Germany & Neustrelitz, Germany & Oberpfaffenhofen, Germany\\
        Min. Satellite Elevation & 10-deg & 20-deg & 5-deg & 5-deg & 5-deg\\
        Height a.s.l. & 489-m & 72-m & 549-m & 66-m & 615m\\
        Primary Aperture & 600-mm & 700-mm & 700-mm & 700-mm & 800-mm\\
        Wavelengths (Rx) [nm] & 1064, 1550 & 1064, 1550 & 1064, 1550 & 1064, 1550 & 589, 850, 1064, 1550\\
        Wavelengths (Tx) [nm] & 1530, 1590 & 1590 & 1590 & 1590 & 1064, 1590\\
        \textit{Standard Compatibility} & CCSDS O3K, SDA & CCSDS O3K & CCSDS O3K & CCSDS O3K & CCSDS O3K\\
        Mean Availability & 65\% & 25\% & 34\% & 32\% & 34\% \\
        \bottomrule
    \end{tabular}
    \label{tab:ogsfeatures}
\end{table}

A cloud-coverage analysis was performed considering the German OGSN and the EONN locations, using the ONUBLA+ software tool. This software exploits a database of weather satellite measurements to determine the cloud-free availability of a selected OGS location. This can be scaled up by selecting multiple locations to create an OGS network, and retrieve the network-wide cloud-free availability. This metric is defined as the percentage of the simulation time over which at least one OGS location has cloud-free visibility. The German OGSN and the EONN networks were created in ONUBLA+, and their cloud-free statistics was investigated using a weather database generated from Meteosat Second Generation satellite measurements from 2009 to 2017. In the weather database, not all exact OGS locations were available. In such cases, the closest location available was selected. Specifically, for the German OGSN, only Oberpfaffenhofen and Neustrelitz locations were available. The other locations were modified as: Munich instead of Neubiberg, Munster instead of Trauen, Granada instead of Almeria. For the EONN, all locations were available, with the exception of Almeria, substituted with Granada for the German OGSN.

\begin{figure}[!htb]
    \centering
    \includegraphics[width=0.6\linewidth]{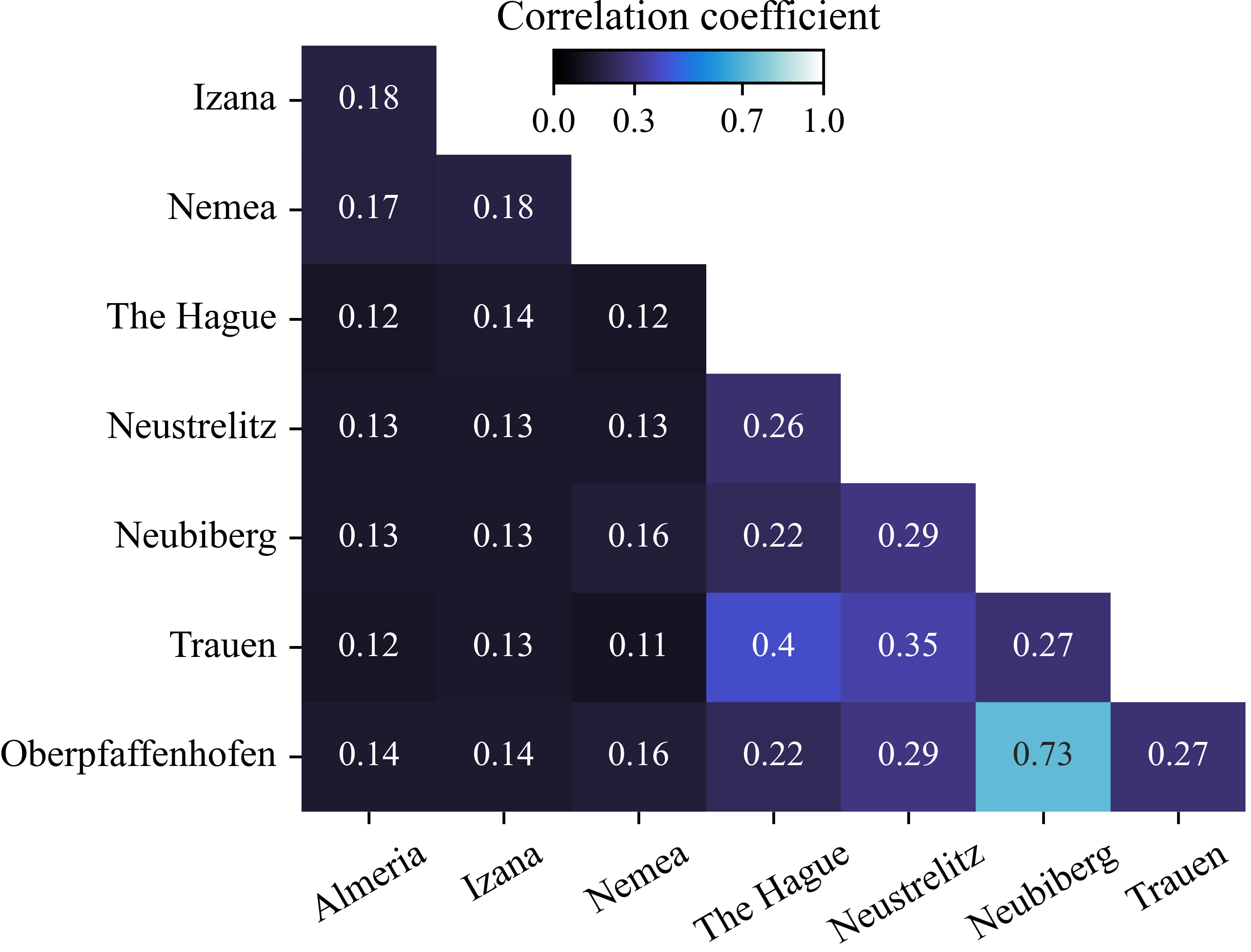}
    \caption{Cloud fraction correlation matrix for the German and EONN locations, based on the ONUBLA+ database.}
    \label{fig:GOGSNCorrelation}
\end{figure}

Figure~\ref{fig:GOGSNCorrelation} presents the cloud fraction correlation values for every OGSN and EONN location combination. It can be observed how the OGSN locations, besides Almeria, show a higher correlation between each other themselves with respect to EONN locations. Correlations among locations belonging to different networks show the lowest correlation values.

\begin{figure}[!htb]
    \centering
    \includegraphics[width=0.5\linewidth]{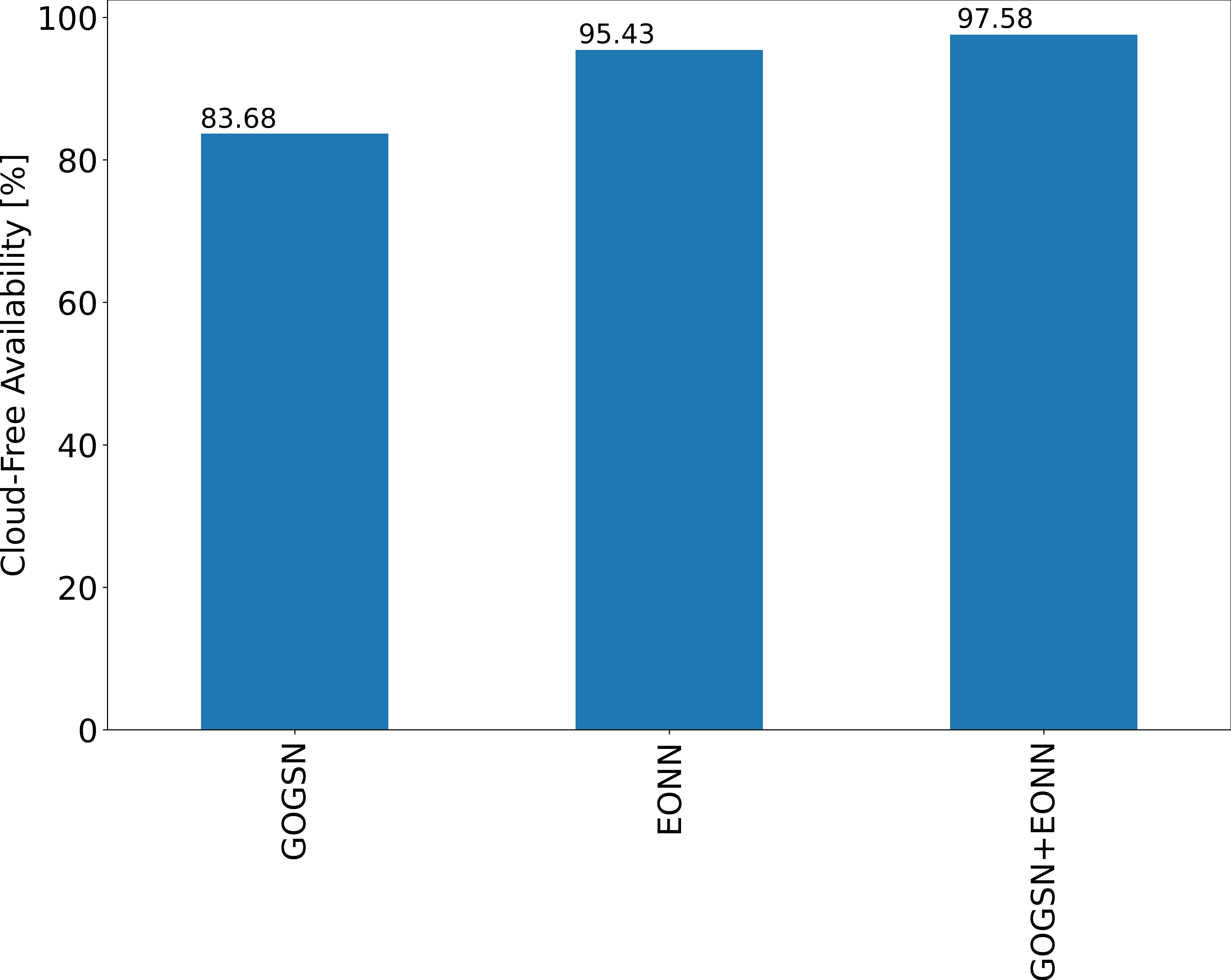}
    \caption{Cloud-free availability for different network configurations in Europe. This value quantifies the proportion of time during which at least one OGS has clear-sky visibility.}
    \label{fig:GOGSNAvailability}
\end{figure}

In Fig.~\ref{fig:GOGSNAvailability}, the network-wide availabilities are shown for single and combined networks. The German OGSN shows only a 83.68\% cloud-free availability, while EONN shows a higher 95.43\% availability. Despite having more OGS sites, the GOGSN has a higher correlation among locations, and German locations do not have the most favourable weather. The EONN network has more locations in Mediterranean areas, and a wide-spread configuration.

\subsection{Small Optical Ground Stations Focal-Optics Assembly: SOFA}
\label{sec:SOFA}

\begin{figure}[!htb]
    \centering
    \includegraphics[width=0.5\linewidth]{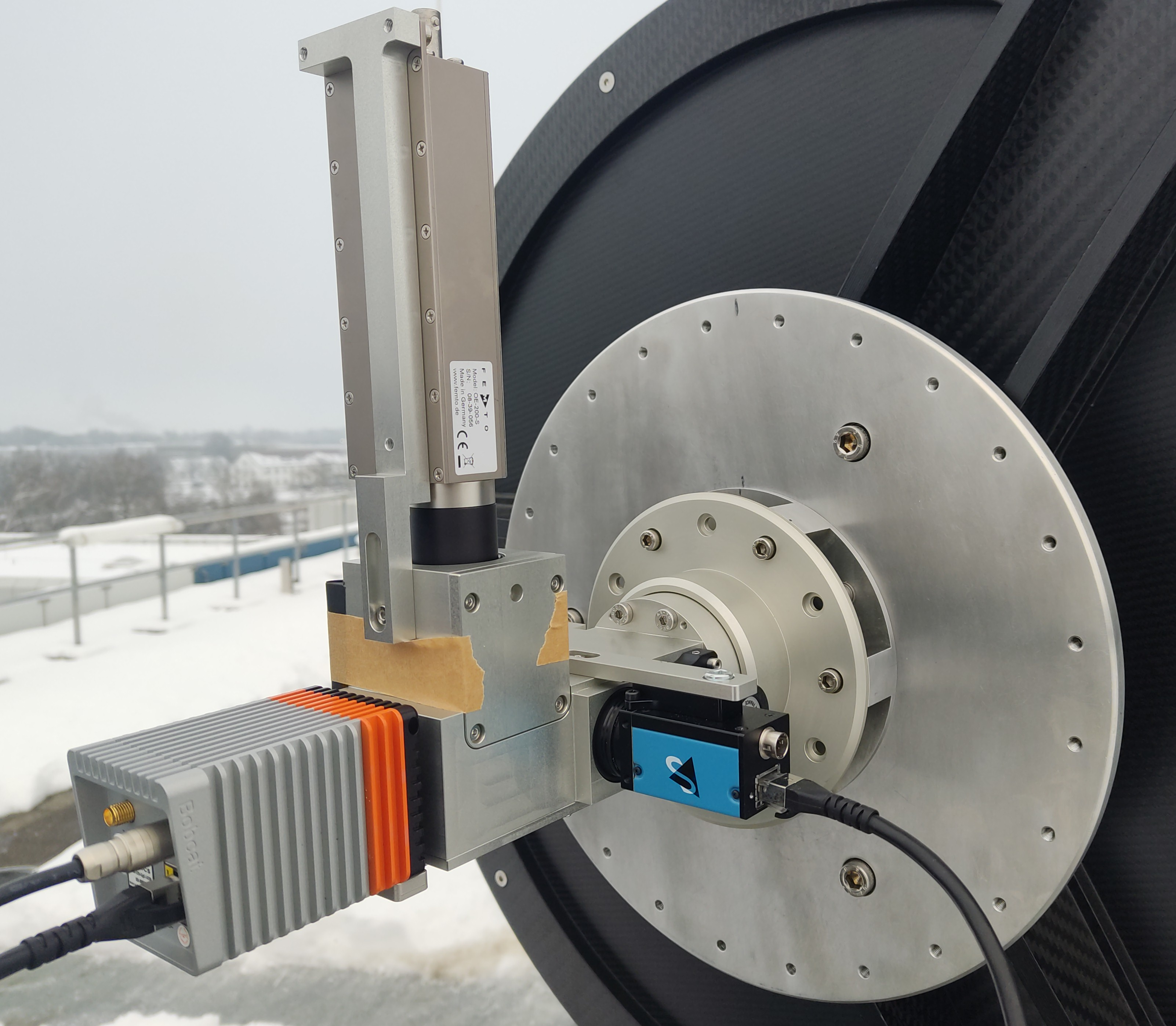}
    \caption{The Small Optical Ground Station Focal-Optics Assembly (SOFA) as mounted on DLR's FOGATA telescope.}
    \label{fig:sofa}
\end{figure}

The DLR's Small Optical Ground Stations Focal-Optics Assembly (SOFA)~\cite{knopp2022}, shown in Fig.~\ref{fig:sofa}, is a small, fully-integrated, low-cost instrument that is to be mounted on COTS or even existing astronomical telescopes enabling optical communication capabilities. As such, a cost-effective and easy deployment of a fully functional optical ground station can be achieved. With its two low-resolution cameras (one each for infrared and day light) the SOFA unit is able to provide all the required information to the telescope mount to achieve a highly accurate open-loop pointing as well as closed-loop tracking of satellites. Detection of the optical signal is realised both with a calibrated InGaAs-PIN photoreceiver for assessment of the optical signal quality and with a free space optical receiver frontend (RFE, small area, fast APD) for data reception. The SOFA unit is equipped with two fiber ports for coupling to an outgoing spatially diverse laser signal to provide a static or modulated beacon to the satellite and for uplink communications.

The SOFA unit is currently deployed on DLR's OGS FOGATA (Almería, Spain) as well as on DLR's OGS LaBoT (Trauen, Germany). Further deployments are planned to the OGS of the University of Auckland, to the OGS of the University of the Bundeswehr Munich (Neubiberg, Germany), as well as DLR's OGSOP (Oberpfaffenhofen, Germany).

\section{Australasian Optical Ground Station Network: AOGSN}
\label{sec:Oz}

The Australasian Optical Ground Station Network (AOGSN) is a partnership between the Australian National University, the University of Western Australia and the Commonwealth of Australia represented by the Defence Science and Technology Group of the Department of Defence, and The University of Auckland, New Zealand. The AOGSN is made up of existing and proposed research infrastructure including optical ground stations and advanced instrumentation technology to demonstrate networking capability across Australia and New Zealand~\citep{Bennet2018}. This section describes the current node capabilities in Australia and site optimisation for node establishment in New Zealand, then presents an analysis of the combined network.

\subsection{Australian National University}
\label{sec:ANU}

The ANU Quantum Optical Ground Station (QOGS) is a 70~\si{\centi\meter} telescope designed for high-performance optical communication. With a small (<25\%) central obscuration and no corrector optics the telescope is optimised for high-throughput of visible and infra-red optical signals. QOGS is situated at the Mount Stromlo Observatory in Canberra, Australia in a purpose-built facility including large optics lab fed via Coud\'{e} path, in addition to Nasmyth foci for optical instrument hosting. The Coud\'{e} room can contain four large optical benches each addressable by optics individually, providing a stable and consistent environment to host both experimental optical instruments and established infrastructure. Multiple locations for optical benches enables equipment from partners, collaborators, and experimental infrastructure under development to access the telescope in an optimum fashion, with only a steering mirror needed to switch between instruments. The telescope is a high-performance PlaneWave Instruments RC700~\cite{satellitetoday} capable of tracking satellites in LEO with a minimum keyhole, even when fully loaded with additional payloads (cite Lafon SPIE2023 124130V). ANU has demonstrated adaptive optics enhanced quantum communication over a horizontal free-space link, and developed adaptive optics for FSOC including satellite-to-ground links~\cite{ martinez2023enhanced}. 

\begin{figure}[!htb]
\begin{center}
    \includegraphics[width=0.9\textwidth]{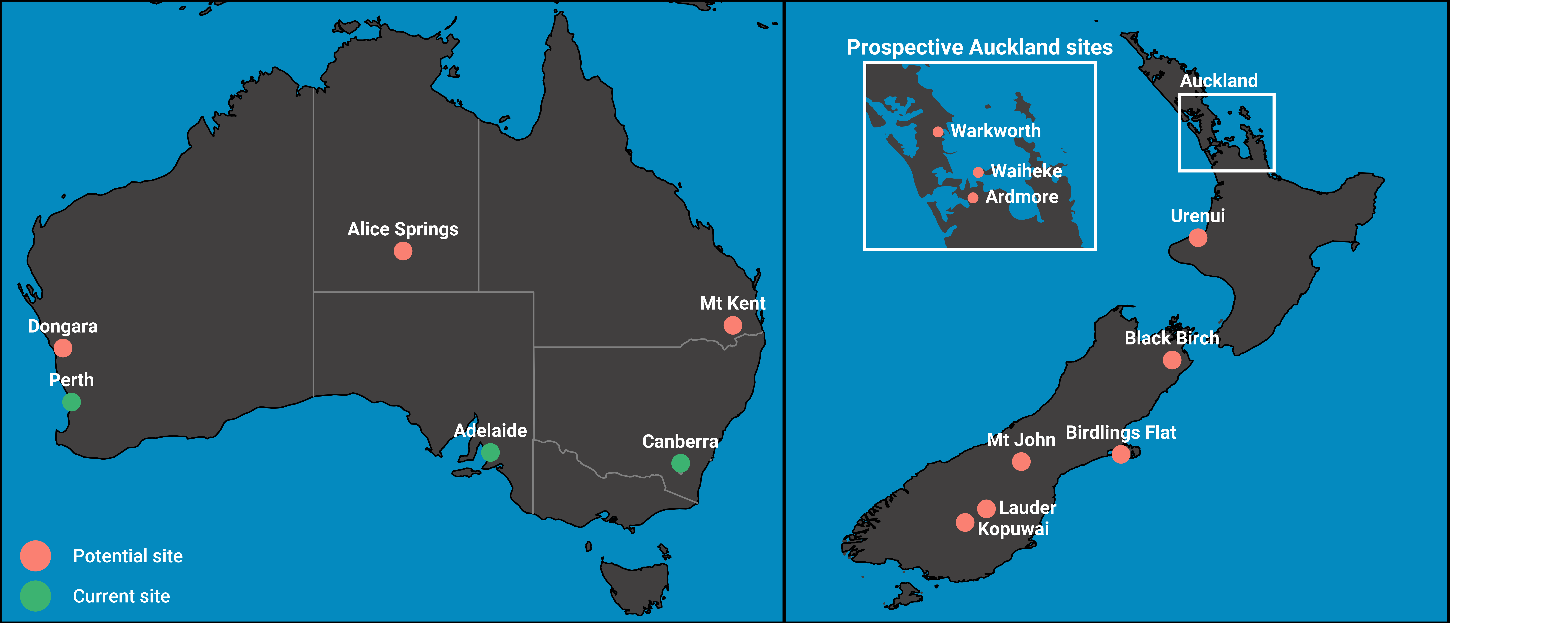}
\end{center}
\caption{The proposed Australasian Optical Ground Station Network~\cite{birch2023availability,Bennet2020}, with nodes across Australia and New Zealand. Green dots indicate current OGS while red dots indicate potential sites.}
\label{fig:AOGSN}
\end{figure}

\subsection{University of Western Australia}
\label{sec:UWA}
The University of Western Australia (UWA) is constructing TeraNet, a three-node optical ground station network within the larger Australasian network. Centred in Western Australia, the aims of the TeraNet project, funded by the Australian Space Agency's Moon to Mars Demonstrator Mission program, are to deploy the technologies needed to realise the potential of optical communications between ground and space, and to develop a transportable ground station.

TeraNet builds on the success of UWA's demonstrations of free-space optical links for metrology and communications ~\cite{p2p_freq, gozzard, dix-matthews_velocimetry, karpathakis_10k, dix_matthews_geo}. Most notably, through a SmartSat CRC funded project a robust single-mode fibre coupling between a deployable optical terminal on the ground and a drone-mounted retroreflector to establish and maintain a 100~\si{\giga\bit\per\second} dual-polarisation quadrature phase shift keying (DP-QPSK) link across 700~\si{\meter} (1.4~\si{\kilo\meter} folded length) at LEO tracking rates \cite{walsh_drone} was demonstrated, followed up with links to both a fixed wing aircraft and a helicopter at 5~\si{\kilo\meter} line-of-sight distance (10~~\si{\kilo\meter} folded, McCann et al. in prep). Through the Australian Space Agency's Moon to Mars Feasibility Mission, a NASA-O2O compatible high photon efficiency link to a drone was demonstrated~\cite{karpathakis_drone}.

TeraNet-1, previously dubbed the Western Australian Optical Ground Station~\cite{walsh2021western}, is a 0.7~\si{\meter} PlaneWave Instruments CDK700 installed on the roof of the physics building at the UWA campus in Perth. It serves as the main research platform where the existing infrastructure and convenient location allow for rapid testing and innovation. First downlink is expected in Q1 2024. TeraNet-2 will be the centerpiece of TeraNet; it will be based on a PlaneWave Instruments RC700 and located at the Mingenew Space Precinct approximately 340~\si{\kilo\meter} north of Perth. This site experiences one of the lowest fractions of cloud cover in the world~\cite{birch2023availability} and is home to NASA's MOBLAS-5 Laser Ranging Station, the most productive station in the International Laser Ranging Service network. 

\begin{figure}[!htb]
\begin{center}
    \includegraphics[width=0.6\linewidth]{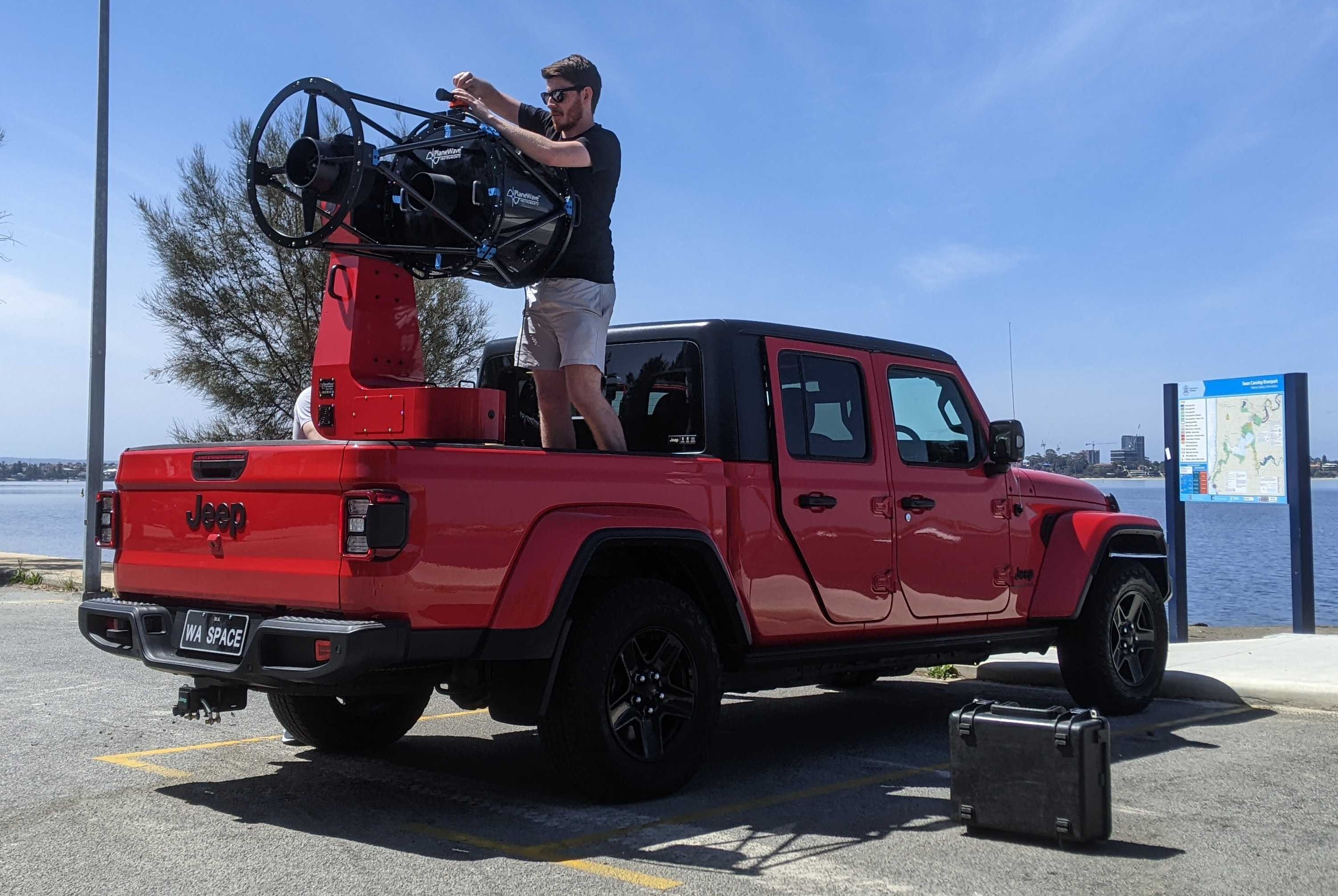}
\end{center}
    \caption{The TeraNet-3 mobile optical ground station.}
    \label{fig:TeraNet}
\end{figure}

TeraNet-3 is a deployable ground station consisting of a PlaneWave instruments CDK17 (0.4~\si{\meter} aperture) mounted on the back of a dual-cab utility vehicle (Figure~\ref{fig:TeraNet}). It is being designed for rapid deployment to demonstrate the benefits optical communications can bring to remote/transient situations such as mine sites, disaster zones, or forward defence deployments. First downlink commissioning is planned for 2024 and will be conducted at ESA's New Norcia Station. In addition to serving as a demonstrator for a rapidly deployable ground station, TeraNet-3 will also serve as a vehicle for outreach activities, both figuratively and literally. 

All three TeraNet nodes will be equipped for direct detection optical communications to and from LEO, with DLR providing access to its OSIRIS payloads for commissioning. Thales will also provide access to its Optel-mu terminals, subject to availability. TeraNet-2 will additionally be equipped with high-order adaptive optics for more advanced capabilities: coherent communications compatibility, optical timing and positioning, and HPE lunar communications. By the end of the Moon to Mars Demonstrator Mission, TeraNet will transition to commercial operations.

\subsection{Defence Science and Technology Group, South Australia}
\label{sec:DSTG}

The DSTG optical ground station, currently under development, is located in Adelaide, South Australia. A major focus of activities is on the assessment and investigation of mitigation strategies for the detrimental effects of the Earth's atmosphere on optical satellite communications. The research emphasis is on sites at low altitudes, which may be desirable locations from an operational perspective but where the effects of atmospheric turbulence are more significant.

The optical telescope is a 0.5~\si{m} diameter Dall-Kirkham telescope with gold coated mirrors and an L-600 tracking mount manufactured by PlaneWave Instruments. Single mode fibre collimators, to be used for the uplink transmitters, are currently under development and will be mounted on the sides of the main telescope. For initial laser downlink demonstrations, an optical communications payload assembly has been constructed, comprising a  beam-splitter, an InGaAs camera, and an avalanche photodetector based communications receiver. The camera, interfaced to the beam-splitter transmit port, is used for alignment and link acquisition and tracking. The communications receiver is interfaced to the beam-splitter reflected port. The assembly is mounted at the telescope’s focus using a lens relay in each arm, with focal lengths selected to adjust the respective FOV’s for the acquisition camera and the photodetector. In future the telescope will be augmented with an adaptive optics payload developed in collaboration with ANU. 

\subsection{New Zealand Optical Ground Station Sites}
\label{sec:NZ}

Establishing a FSOC ground terminal in New Zealand, suitable for inclusion in an international network of FSOC nodes, was the subject of a feasibility study in 2021. This study was in part motivated by a shared desire for closer collaborative research links between New Zealand, Australia and the DLR.

The set of initial sites in the feasibility study comprises locations which range from minimally developed through to a fully provisioned astronomical observatory. The range of sites considered was deliberately wide so as to illuminate the advantages and challenges associated with establishing an OGS at sites at different locations and in varying states of readiness. A brief summary follows of those sites assessed in the feasibility study and which were integrated with ANU studies on cloud cover fraction statistics.

The University of Auckland  maintains a field station at \textbf{Ardmore}, in the north of the North Island. This has been the site of a number of radio physics and atmospheric experiments in the past, and is currently home to a number of environmental science projects. The site is flat and at low altitude. It has power, communications, utility services and a basic mechanical workshop facility. 
\textbf{Urenui} is a settlement on the east coast of the North Island of New Zealand. It is home to the M\={a}ori iwi (tribe) Ng\={a}ti Mutunga. It lies in a region of lower than average cloud fraction and the iwi facilities were upgraded in 2020 as part of the then Government's Provincial Growth Fund~\cite{PGF}.

\begin{figure}[!htb]
\begin{center}
    \includegraphics[width=0.4\linewidth]{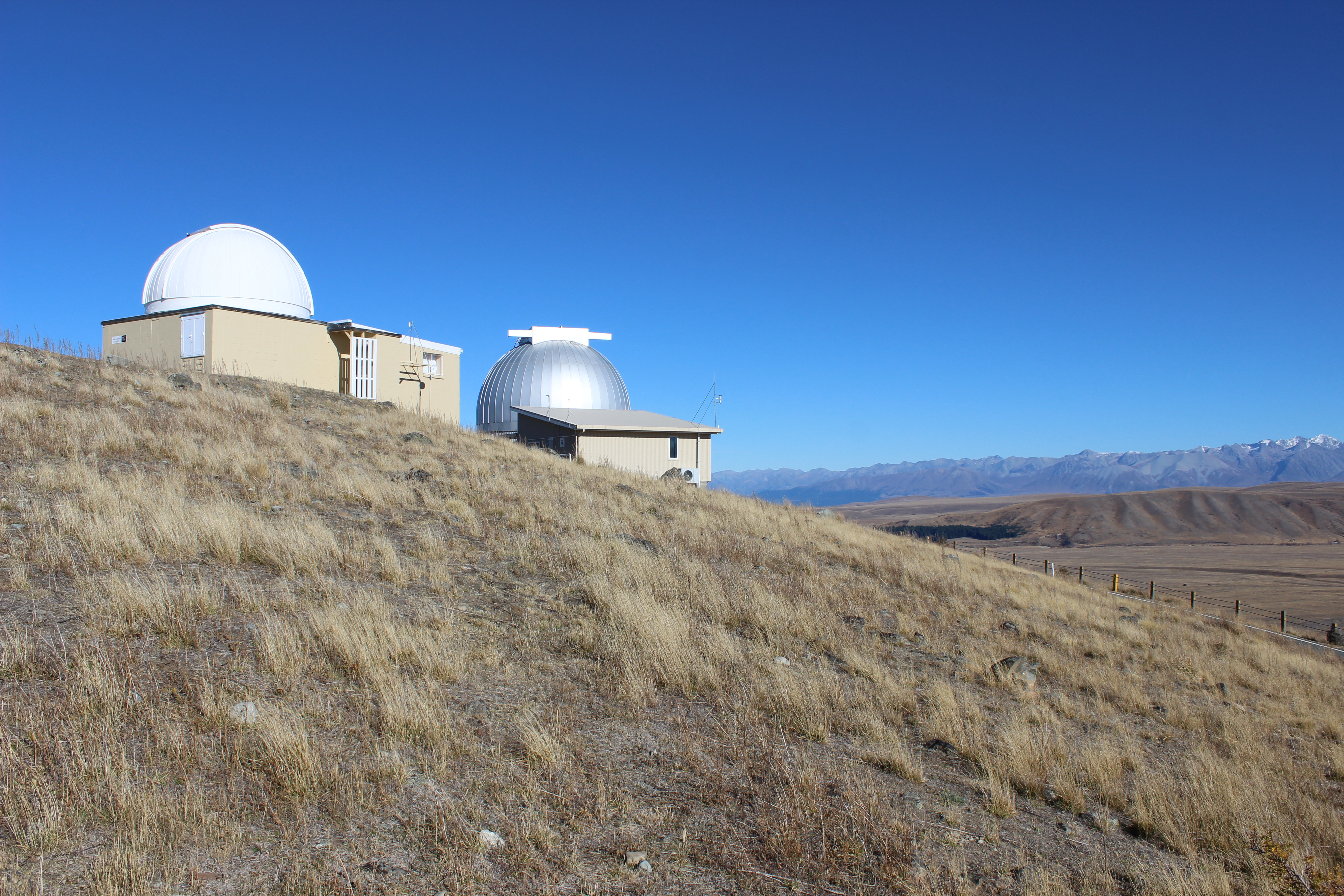} 
    \includegraphics[width=0.4\linewidth]{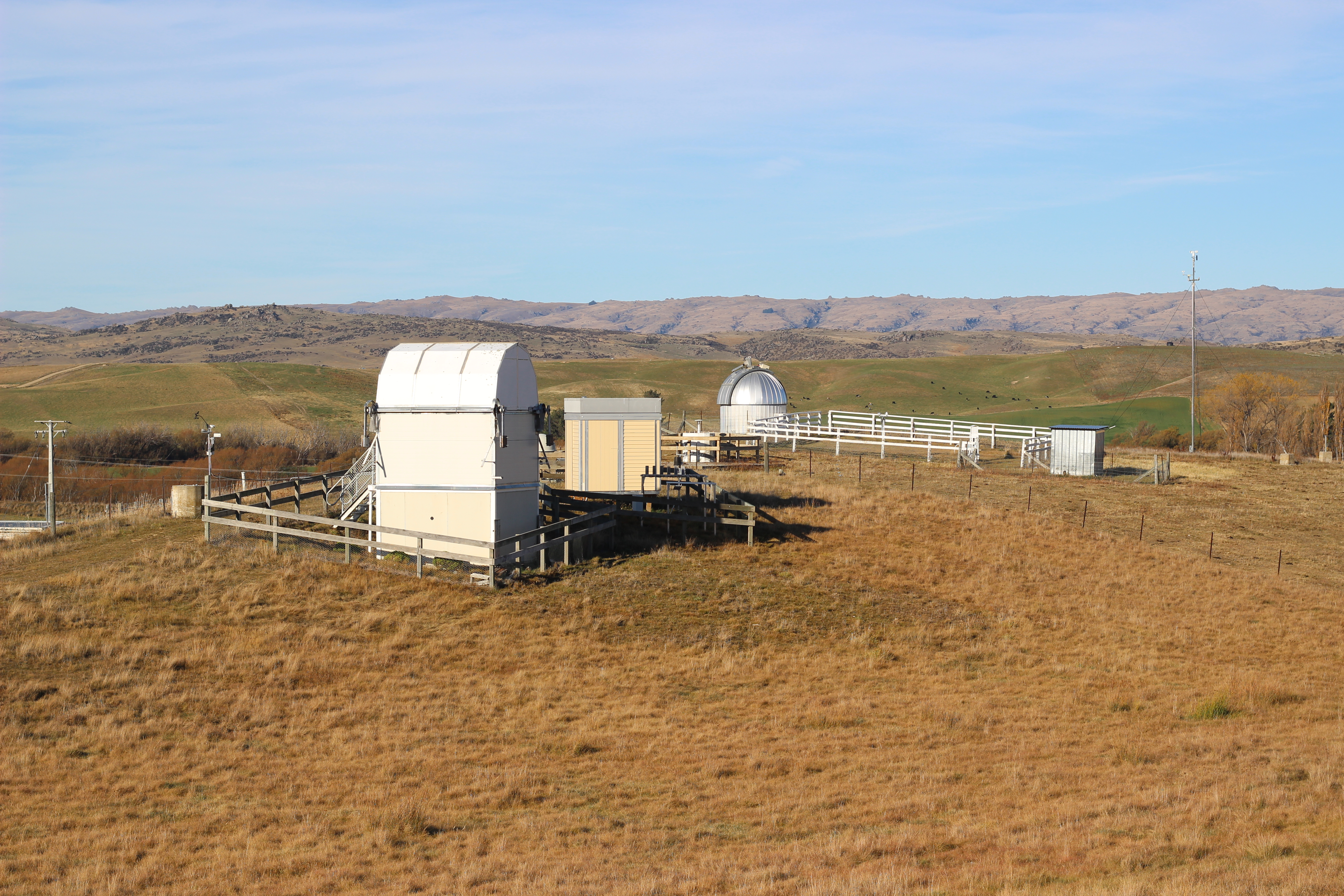}
\end{center}
\caption{Mount John Observatory domes (left) housing the 0.61~\si{\meter} B\&C telescope and the 1.8~\si{\meter} Nishimura telescope. Sensing experiments at NIWA's Atmospheric Research Station, Lauder. The 0.6~\si{\meter} Yock-Allen BOOTES-III robotic telescope enclosure is in the foreground (right).}
\label{fig:mjo1}
\end{figure}

In the South Island, the \textbf{Black Birch} Observatory site was used as a centre of operations in an earlier site testing programme for an astronomical observatory\cite{bateson}. It was also the southern sky observatory of the US Naval Observatory, operating from around 1982 to around 1995. The facilities on the site were dismantled and removed in 1997 by amateur astronomers and the site was returned to nature.
The \textbf{Mount John Observatory} is the home of four professional telescopes: the 1.8~\si{\meter} Nishimura telescope, the 1~\si{\meter} McLellan telescope, the 0.61~\si{\meter} Optical Craftsman and the 0.61~\si{\meter} Boller and Chivens telescope, see Fig.~\ref{fig:mjo1}. Road access is excellent, mechanical and electrical engineering facilities are available onsite, and there is accommodation for extended stays by observers and staff. \textbf{Birdlings Flat} is an area 40~\si{\kilo\meter} south of Christchurch and has been the location of space-related activities, including launch. Nearby is the the Kaitorete spit, which is the location of the T\={a}whaki Joint Venture, a commercial partnership between two r\={u}nanga (a r\={u}nanga is an authority set up by and for a tribe) and the New Zealand Government. 

Further south, NIWA operates an Atmospheric Research Station (ARS) at \textbf{Lauder} in Central Otago. The site specialises in measuring CFCs, ozone, UV levels and greenhouse gases. The NIWA ARS at Lauder has an extensive array of atmospheric sensing equipment, including spectrometers, radiometers, all-sky cameras and clear-sky detectors. \textbf{Kopuwai/Obelisk} is a natural stone feature atop a mountain peak in the Old Man Range, in Otago. The peak is accessed by an unsealed road which is one of the highest in the country.Finally, at the southern tip of the South Island, the \textbf{Awarua} satellite ground station (SGS) is located approximately 10~\si{\kilo\meter} from Invercargill, New Zealand's southernmost and most westerly city. 

\begin{figure}[!htb]
\begin{center}
     \includegraphics[width=0.4\linewidth]{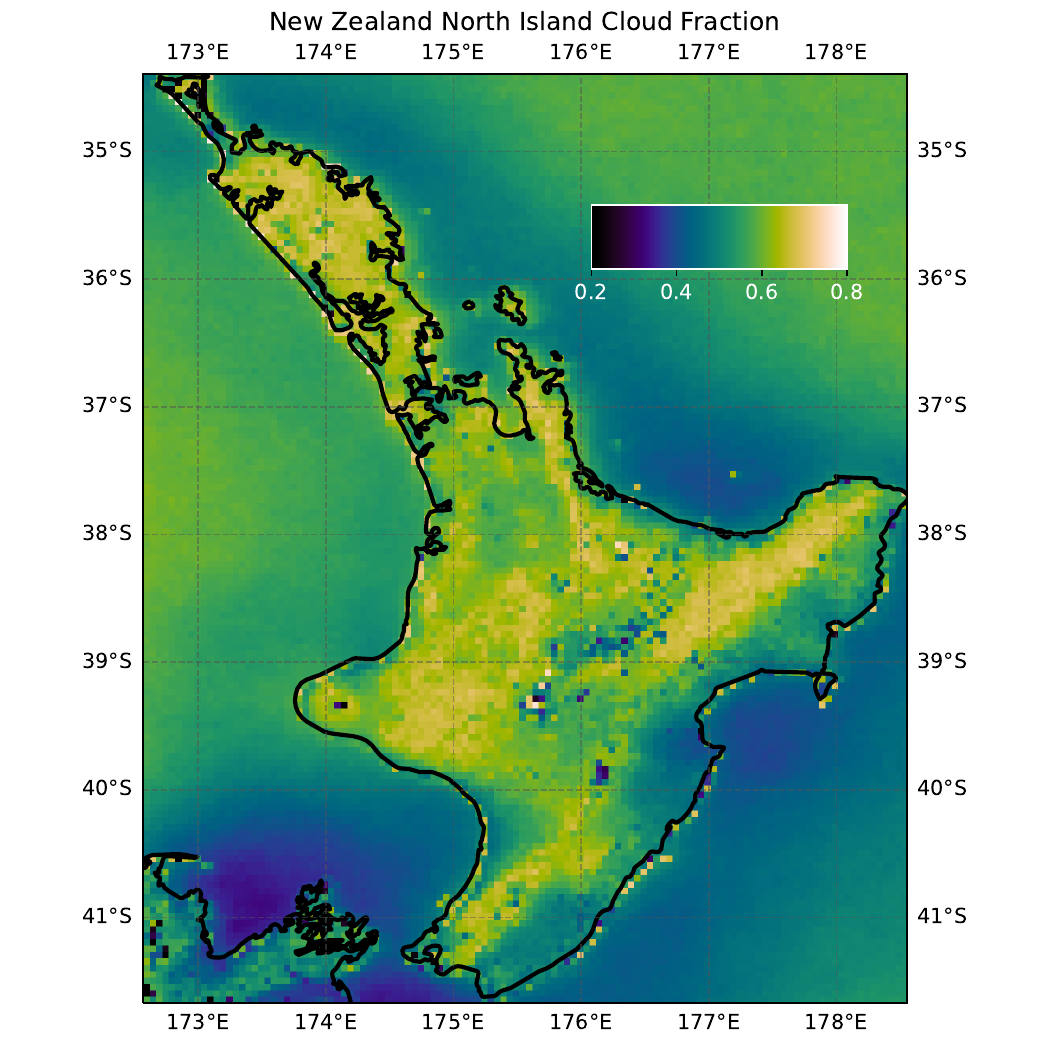} 
     \includegraphics[width=0.4\linewidth]{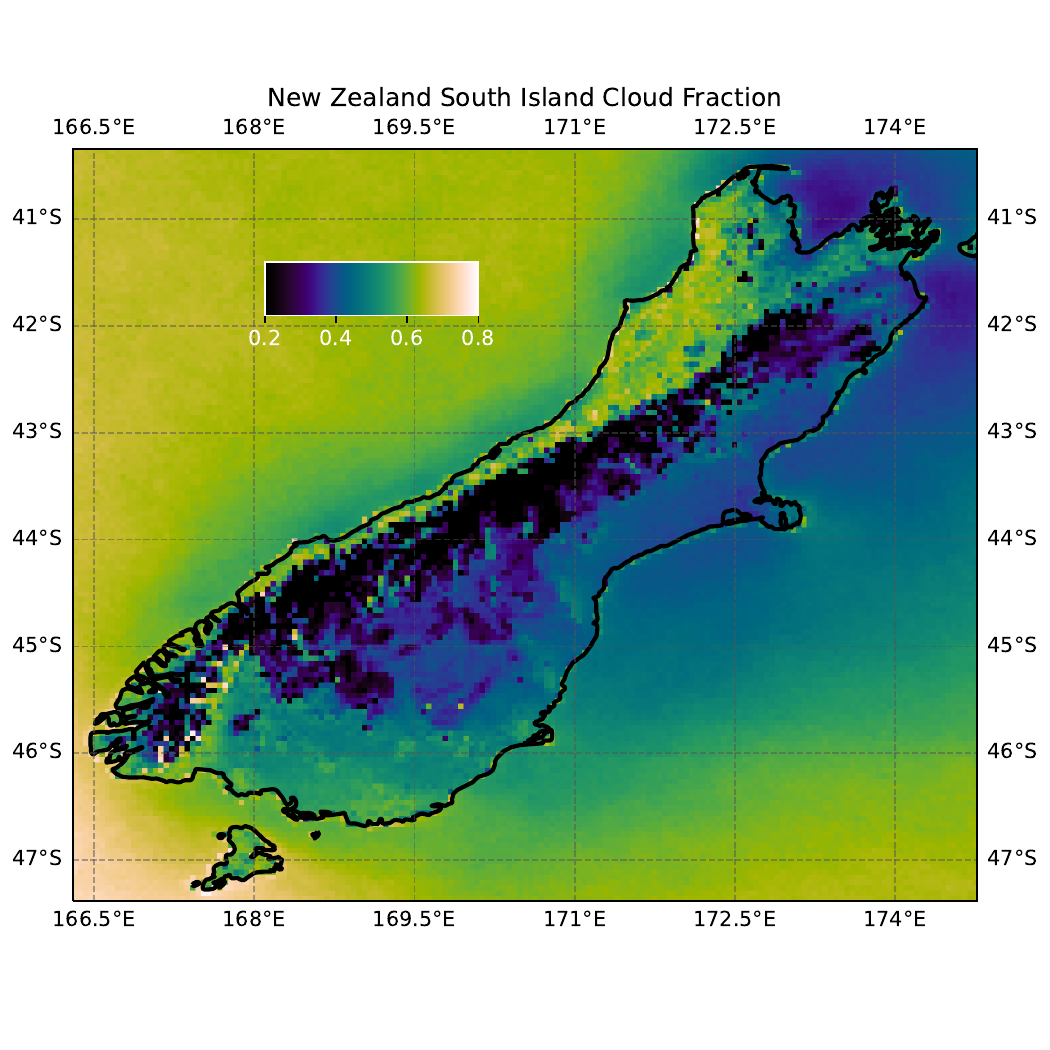}
  \label{fig:cloudfraction}
\end{center}
\caption{Cloud coverage fraction for New Zealand, showing the North Island (left), and South Island (right).}
\label{fig:cloudfrac}
\end{figure}

Figure~\ref{fig:cloudfrac} shows a plot of cloud fraction for New Zealand. It is observed that the lowest values occur in the mountainous regions in the South Island of the country.

Identifying more than one potential site in New Zealand for FSOC is important as the nature of the country's geography results in weather patterns that are highly variable both on spatial and temporal scales. Therefore, as an outcome of the initial feasibility study, further sites were added. These sites are \textbf{Waiheke Island}, and the satellite communication station north of \textbf{Warkworth}. Waiheke Island in the Hauraki Gulf is relatively near to Auckland and enjoys a sunny micro-climate that supports a vineyard operated by The University of Auckland. Operations at the satellite communication station in Warkworth have recently been transferred to Space Operations New Zealand, which operates a satellite communications station at Awarua. 

\begin{figure}[!htb]
\begin{center}
    \includegraphics[width=0.6\linewidth]{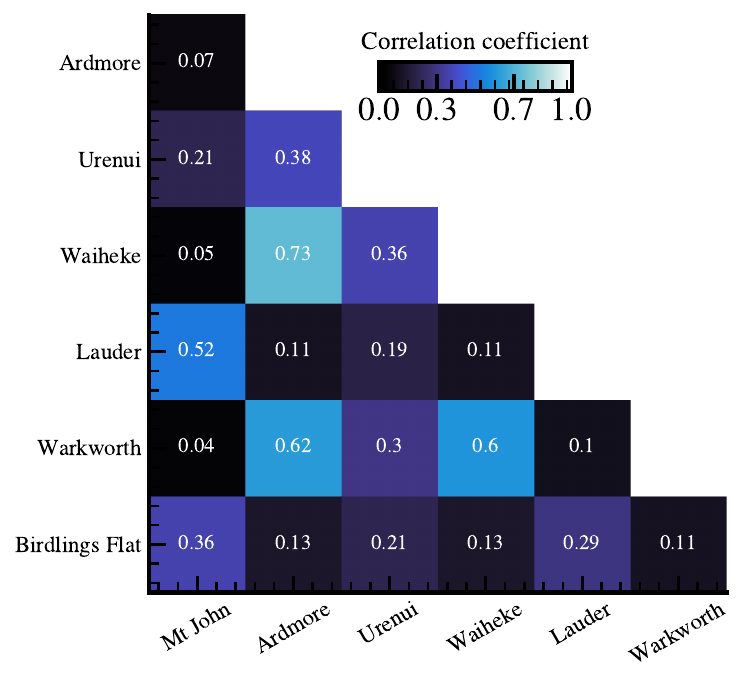}
\end{center}
    \caption{Cloud fraction correlation matrix for prospective New Zealand OGS locations. Measurements with AHI/\textit{Himawari-8} from 2015-2022.}
    \label{fig:corr_matrix}
\end{figure}

A site diversity analysis for a selection of potential sites in New Zealand was performed and the results are shown in Fig.~\ref{fig:corr_matrix}. The spatial correlation analysis used data obtained during 2015-2022 using the Advanced Himawari Imager (AHI) on Himawari-8. Two sites, \textbf{Black Birch} and \textbf{Kopuwai}, are shown in Fig.~\ref{fig:AOGSN} but excluded from further analysis. The \textbf{Awarua} site was also considered in the feasibility study but was similarly excluded from the site selection.

Sites which are geographically close, such as the Auckland sites of Ardmore and Waiheke, show relatively high correlation values as expected from sharing similar weather patterns. These data are useful when considering where a second ground station node could be built in New Zealand, based on the location of a first ground station. The addition of a second (or further) ground station nodes will improve the reliability of the NZ contribution to the Australasian network. A complete three-site diversity correlation analysis of NZ nodes remains for future work.

In order to accelerate activity and capability in FSOC operations, New Zealand researchers have developed a prototype FSOC ground station at the Ardmore site. A 10$'$ observatory dome (PD10, Technical Innovations) has been installed and houses an L-350 mount from PlaneWave Instruments, Inc. The intended payload comprises the DLR SOFA (Small Optical Ground Stations Focal-optics Assembly) unit~\cite{knopp2022} and an RC telescope with a 25~\si{\centi\meter} primary. Installation of the OTA and SOFA unit is planned for Q1 2024. For calibration and early testing purposes, a secondary telescope (12$''$ LX200GPS, Meade Instruments) has been mounted on the L-350. A set of three all-sky cameras, developed by NIWA, have been deployed at various sites, including one at Ardmore. These cameras generate cloud image data which is being used in current research into short-term (i.e., on the time-scale of a typical LEO satellite pass) cloud coverage forecasting. Finally, in order to progressively characterise the Ardmore site as an OGS, an integrated seeing monitor (Miratlas) was installed, which comprises a turbulence monitor and weather station. 

\subsection{Combined Australian-New Zealand FSOC network Performance}
\label{sec:Combined}

\begin{figure}[!htb]
\begin{center}
    \includegraphics[width=0.9\linewidth]{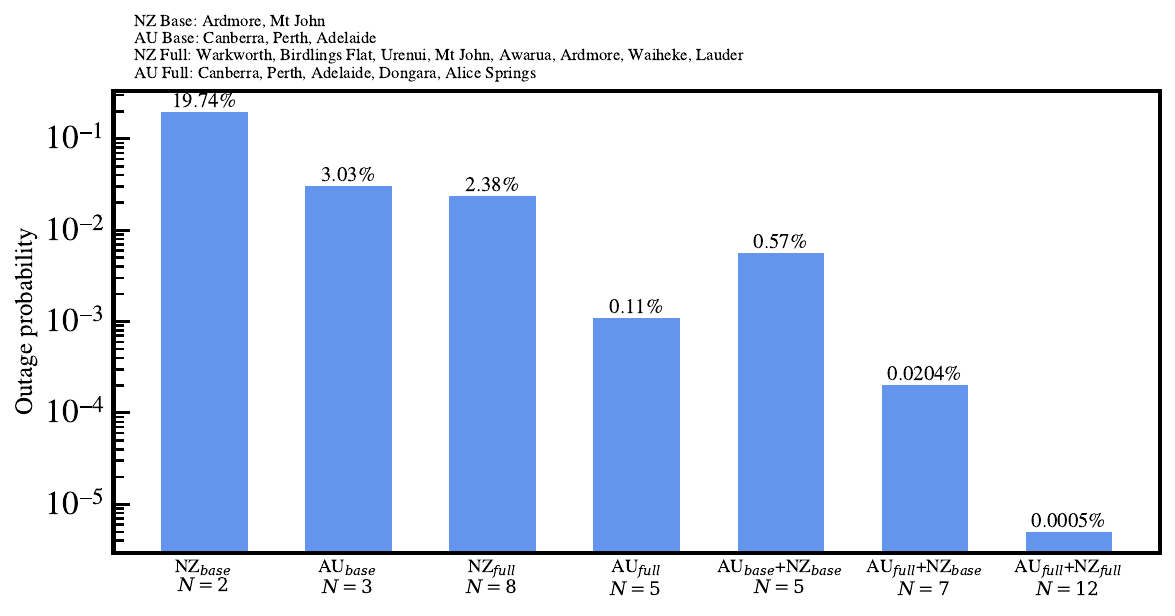}
\end{center}
    \caption{Outage probability for numerous AOGSN network configurations. This is defined as the probability that all nodes in a network simultaneously suffer an outage due to cloud fraction and is computed using the spatially-correlated Bernoulli model~\cite{birch2023availability}.}
    \label{fig:outage_probs}
\end{figure}

A reliability analysis has been performed to optimise the AOGSN  configurations and highlight the improvements gained from adding New Zealand OGS to extend existing Australian OGS sites. Networks in Australia and New Zealand have been assumed as base or full variants, with the set of sites corresponding to these configurations in the title of Fig. \ref{fig:outage_probs}. Reliability estimates are made using the spatially-correlated neuron spike model and with cloud fraction data from VIIRS/SNPP and cloud fraction covariances from AHI/\textit{Himawari-8}~\cite{birch2023availability}. In Fig.~\ref{fig:outage_probs} reliability is represented as minimising the outage probability. The nodes used in Fig.~\ref{fig:outage_probs} can be identified in Fig.~\ref{fig:AOGSN}. A particularly important result in Fig. \ref{fig:outage_probs} is the $3\%$ outage probability of the existing 3 OGS nodes in Australia being reduced to $0.57\%$ with the addition of just two OGS in New Zealand, Ardmore and Mt John. Note that this analysis only considers reliability, not capacity or coverage, and that New Zealand significantly expands the longitudinal coverage of a Australasian network.

\section{Future Network Development}
\label{sec:NetDev}

Future research work includes improving the throughput of the optical channel for FSOC, securing it using quantum technologies, and maximising the resilience of FSOC ground segment networks. This section summarises some of the work underway in Germany and Australasia towards these goals. 

\subsection{German FSOC Research}
\label{sec:GFSOCR}

Current research areas associated with the FSOC ground stations network in Germany include measurements of the optical channel to analyze and forecast the availability of optical links in dependence on the geographical location. The results will facilitate updates on existing or the definition of new site diversity and channel models. Basic research on the optical channel is the foundation for further developments of FSOC systems. For example, a precise characterization of the optical channel facilitates the development of new transmission techniques such as multiple-input multiple-output (MIMO) for FSO systems~\cite{Hung2022MIMO}. 
A related field of research is the development of new forward error correction codes to counter atmospheric disturbances.

\subsection{Adaptive Optics} 
\label{sec:AO}

FSOC systems use adaptive optics to improve performance~\cite{Wang_etal:2018, martinez2023enhanced}. The Mount John Observatory (MJO) is one of the leading candidates for an OGS in New Zealand, and as there are existing installed telescopes, work has commenced on developing AO capability at this facility. 
A low-order (tilt) closed-loop adaptive optics system has been installed on the Boller \& Chivens telescope at MJO~\cite{Liu2020}. System design parameters were sourced from earlier work developing a turbulence site profile~\cite{mohr_cottrell:2010}.
This system was originally developed to partially correct for the adverse effects of imaging space debris through the Earth's turbulent atmosphere~\cite{Muruganandan_etal:2023}. Target objects included large, defunct satellites, both in LEO and GEO. More recently, the tilt AO corrector was used to improve the stability of fringe patterns generated by a modified Michelson interferometer used to measure spatial coherence of satellites for space domain awareness. 

The effectiveness of an AO system is highly dependent on source wavelength, where photon absorption (depletion at the sensor or FO collimator) is higher at shorter wavelengths. This suggests that some classical FSOC systems may not require AO, whereas for quantum FSOC systems, which use a shorter wavelength than more conventional systems, a high-order AO system would be considered essential. Currently, a fifth-order dual-mirror (woofer-tweeter) AO system is in development by the Department of Electrical \& Computer Engineering at the University of Canterbury that may meet these requirements.

\subsection{Quantum Secured Communications}
\label{sec:QKD}

Quantum key distribution (QKD) is a method in which communicating parties generate and share a secret key where the presence or information leakage to an eavesdropper can be inferred by fundamental quantum mechanics~\cite{Bennett1984Quantum}. The secret key can then be used to encrypt information sent between parties. QKD networks have been developed globally where the distribution of secret keys have predominantly been fibre-based using DVQKD protocols~\cite{Cao2022Evolution}. The most significant demonstration of QKD has been the demonstration of the DVQKD protocol, BB84, from the Micius satellite which distributed secret keys between two optical ground stations in Xinglong, China and Graz, Vienna, 7,600~\si{\kilo\meter} apart~\cite{Chen2021Integrated}. The demonstration achieved a sifted key rate between 3--9~\si{\kilo\bit\per\second} in good weather conditions between a free-space link distance of approximately 1000~\si{\kilo\meter} and 600~\si{\kilo\meter}, respectively, validating the feasibility of QKD between LEO satellites and optical ground stations.

In Europe, the European Quantum Communication Infrastructure (EuroQCI) Initiative will make use of innovative quantum communication technologies that are developed, for example, in EU-funded projects such as the Horizon 2020 OpenQKD project~\cite{OpenQKD} or the Quantum Technologies Flagship~\cite{QuantumFlagship}. It will be, furthermore, an integral part of IRIS$^2$, the new EU space-based secure satellite communication system~\cite{IRIS2}. 

Each optical ground station discussed in this work will in principle be compatible with QKD though a ground stations suitability for a particular mission would need to be assessed based on the compatibility of transmit and receive apertures. For example a 30~\si{\centi\meter} transmission aperture on a LEO satellite would require a $\simeq$1~\si{\meter} receive aperture to achieve a positive secret key rate using continuous variable (CV) QKD ~\cite{Sayat2024Satellitetoground, Kish2020Feasibility}. 

\subsection{Network Expansion}
\label{sec:NetworkExpansion}

The utility of each ground station could be extended by connecting to wider ground based fibre networks to link a satellite with multiple ground locations near each optical ground station. Moreover, the distributed network of OGSs constitutes an excellent test-bed to showcase new application scenarios and demonstrate new hard- and software developments.

A natural expansion of the proposed network would be to include the Antarctic continent in order to provide the continent with a high-bandwidth solution that it currently lacks. Australia and New-Zealand are active in research in Antarctica, occupying  coastal locations including Casey, Mawson Davis, and Scott bases. These provide possible locations to start expanding the network to the Antarctic region thanks to the existing logistics and relative ease of access. The geographic location of the continent is advantageous to this technology~\citep{Bennet2020} as polar-crossing orbits are favoured by many satellite operators, and therefore there could be many overhead passes of a polar station per day. The interior of the continent, where internet access is the most challenging, has ideal weather conditions for such technology. The skies are amongst the clearest in the world, with 20~$\%$ cloud cover~\citep{cloud}. There is also ample evidence that seeing conditions are excellent which attracts astronomers to the continent~\citep{lawrence2004exceptional,aristidi2005analysis,aristidi2005site}. Even during the daytime, which can last for months, there are periods where the seeing drops to less than 1'' due to the thermal balance between the atmosphere and the highly radiating ice. 
 
To establish a truly global optical communication ground station network, the German and Australasian networks could be combined. Success of such an endeavor will require the continued exchange of best practices and knowledge, and progressively increasing coordination between each network.

\section{Conclusions}
\label{sec:conclusions}
Successfully and routinely communicating between Earth and space using laser light demands the establishment of a network of interconnected ground stations, located so as to maximise the number of opportunities to establish a link with spacecraft. The performance of a network depends on the geographical disposition of the network nodes and the weather conditions typically experienced at each node. 

Spacing ground stations in longitude across the globe provides favourable network access to satellites for typical orbit choices for a telecommunications satellite cluster. Australia and continental Europe are two large land masses extended longitudinally and are developing technologies and operational facilities suitable for inclusion in a optical communications ground station network. The research and commercial opportunities in FSOC has resulted in an increase in funding for this area in Australia and New Zealand and Germany continues its long history of research leadership in FSOC.

Technical developments across the sector have drawn a future FSOC network closer to reality. Modern observatories, comprising telescope mounts capable of tracking satellites reliably are now commonly available on the market and instrumentation such as the DLR's SOFA unit offers a rapid route to establishing a standard COTS observatory as an FSOC ground segment node in a network. 

Site testing is a key ingredient in establishing where optical ground stations might be most favourably placed in an FSOC network. This work presented the two-site correlation values for several locations in New Zealand, derived from an analysis of satellite cloud coverage data. The analysis also included measures of outage probability of a network comprising various combinations of FSOC terminals wholly within each of Australia and New Zealand, and across both nations. The addition of two New Zealand nodes in an Australian network decreased the outage probability of the combined network from $~3\%$ to $~0.6\%$. Owing to the general north-south orientation of New Zealand, additional FSOC terminals will generally only add to the resilience of the network against local weather patterns and not add more grasp of the network that would come from extending the network further in longitude. 

\section*{Acknowledgments}
The authors acknowledge the Japan Meteorological Agency (JMA), which operates the Himawari-8 satellite. NJR acknowledges Te R\={u}nanga o Ng\={a}ti Mutunga. 

\subsection*{Author contributions}

NJR contributed to Sections~\ref{sec:Intro}, \ref{sec:NZ} and \ref{sec:conclusions}. JA contributed to Section \ref{sec:NZ}. TT contributed to Section~\ref{sec:NetworkExpansion}. MB contributed to Sections~\ref{sec:Intro} and \ref{sec:Oz}. JEC contributed to Sections~\ref{sec:NZ} and \ref{sec:NetworkExpansion}. MTS and OT contributed to Section~\ref{sec:QKD}. RTS and AK contributed to Sections~\ref{GOGSN} and \ref{sec:GFSOCR}, KM contributed to Section~\ref{sec:DSTG}. DG, JR, MTK contributed to Sections \ref{sec:German} and \ref{sec:GFSOCR}. SR contributed to Section \ref{GOGSN}.

\subsection*{Financial disclosure}

NJR, JEC, SW, CQ, KW and JA receive funding for this work from New Zealand's Ministry of Business, Innovation and Employment via a Catalyst: Strategic – New Zealand-DLR Joint Research Programme December 2020 and/or a 2021 Catalyst Strategic - New Zealand DLR Joint Research Programme grant. 

The Optical Ground Station Neubiberg, Munich, Germany, is funded by dtec.bw - Digitalization and Technology Research Center of the Bundeswehr. dtec.bw is funded by the European Union - NextGenerationEU.

The Quantum Optical Communication Ground Station is funded by A PIP grant from the ACT government and supported by the Moon to Mars Initiative. 

The TeraNet project is funded by the Australian Space Agency's Moon to Mars Demonstrator Mission, the Government of Western Australia, and the University of Western Australia.

The Laser Ground Station Trauen (LaBoT) is supported by the German Federal Ministry of Defense through the technological research and development assignment ``Responsive Space Capabilities'', the infrastructure has been subsidized by the European Regional Development Fund (ERDF). 

\subsection*{Conflict of interest}
The authors declare no potential conflict of interests.

\bibliography{FSOC_Bibliography}

\end{document}